\documentclass[
    ,final            
]
  {aipproc}
\usepackage{bm}
\layoutstyle{6x9}

\newcommand{\tr}{{\textrm {tr}}}

\newcommand{\Tr}{{\textrm {Tr}}}

\begin{document}

\title{Non-perturbative contributions to the Polyakov loop above the deconfinement phase transition\thanks{Presented by E. Meg\'{\i}as at the Quark Matter 2005, Budapest, Hungary, August 4 to 9, 2005.}}

\author{E. Meg\'{\i}as, E. Ruiz Arriola and L.L. Salcedo}
{address={Departamento de F{\'{\i}}sica Moderna,  
Universidad de Granada. 18071-Granada (Spain)}
}

\begin{abstract}
The Polyakov loop is an order parameter to the
confinement-deconfinement phase transition in pure
gluodynamics. Recently a simple phenomenological model has been
proposed to descri\-be the available lattice data for the renormalized
Polyakov loop in the deconfinement phase~\cite{Megias:2005ve}. These
data exhibit unequivocal inverse power temperature corrections driven
by a dimension two gluon condensate. The heavy quark free energy is
also studied and compared to lattice data.
\end{abstract}

\maketitle



{\bf Introduction.} In pure gluodynamic the confinement-deconfinement
transition can be characterized by the breaking of the Z($N_c$)
discrete symmetry, belonging to the center of the gauge group,
SU($N_c$). The order parameter is the traced Polyakov loop, defined by
\begin{equation}
L(T)= \langle \tr \; \Omega(x) \rangle =\Big\langle \frac{1}{N_c} \tr \;{\mathbf P} \left( e^{ig\int_0^{1/T} dx_0 A_0(\mathbf{x},x_0)} \right)\Big\rangle \,,
\label{eq:def_polyakov_loop}
\end{equation}
where $\langle \; \rangle$ denotes vacuum expectation value. $A_0$ is the gluon field in the (Euclidean) time direction. A perturbative evaluation of the Polyakov loop was carried out in~\cite{Gava:1981qd}. 

The Polyakov loop correlation functions are related to the change in
the free energy arising from the presence of a static quark-antiquark
pair in a thermal medium. The colour singlet free energy yields the
heavy quark potential at finite temperature, 
\begin{equation}
F_1(\mathbf{x},T)/T=-\log \langle \Tr  \; L(\mathbf{x}) L^\dagger (0) \rangle + c(T) \,.
\end{equation}

A renormalization procedure for the Polyakov loop based on the
computation of singlet and octet correlation functions between
Polyakov loops in the limit of large separation has been considered in
recent lattice studies \cite{Kaczmarek:2002mc,Kaczmarek:2005ui}.  The
additive normalization constant, $c(T)$, is ambiguous and can be fixed
at short distances by comparison with the zero temperature heavy quark
potential, i.e.  for the singlet free energy: $F_1(r \ll 1/T) \simeq
V_{qq}(r)$. With such a prescription the renormalized Polyakov loop is
defined as~\cite{Kaczmarek:2002mc,Kaczmarek:2005ui}:
\begin{equation}
L^{ren}(T)= \exp \left( -\frac{F_1(r \rightarrow \infty,T)}{2T}\right) \equiv 
\exp \left( -\frac{F_\infty(T)}{2T}\right) \,.
\label{eq:ren_polyakov_loop}
\end{equation}
In Ref.~\cite{Megias:2005ve}, we have proposed a model to describe the
available lattice data for $L^{ren}$. Here we will show that it also
describes consistently the lattice results for the free
energy.

{\bf Phenomenological model for the Polyakov loop.} From now on we
work in the Polyakov gauge: $\partial_0 A_0 (\mathbf{x},x_0)=0$. We
assume that, in the deconfinement phase, the field $A_0(\mathbf{x})$
is sufficiently well described by a Gaussian distribution. From
(\ref{eq:def_polyakov_loop}) one finds~\footnote{The Gaussian ansatz
is correct up to ${\cal O}(g^5)$ in perturbation theory, and exact in
the large $N_c$ limit.}
\begin{equation}
L = \exp \left[ \frac{-g^2 \langle A_{0,a}^2 \rangle_T}{4 N_c T^2}\right] \,.
\end{equation}

To describe the dynamics of the $A_0(\mathbf{x})$ field we use the
3-dimensional reduced effec\-tive theory of
QCD~\cite{Appelquist:1981vg,Megias:2003ui}. Let $D_{00}(\mathbf{k})\delta_{ab}$ denote
the 3-dimensional propagator, then
\begin{equation}
\langle A_{0,a}^2\rangle_T = (N_c^2-1)T \int \frac{d^3k}{(2\pi)^3} D_{00}(\mathbf{k}) \,.
\label{eq:cond}
\end{equation}
In perturbation theory the propagator becomes $D_{00}^P(\mathbf{k}) =
1/(\mathbf{k}^2+m_D^2) $ ($m_D \sim T $ is the Debye mass), and when
inserted in~(\ref{eq:cond}) it reproduces the known perturbative result
to LO~\cite{Gava:1981qd}. We will take into account the
non-perturbative contributions coming from dimension -2
condensates. So, we consider adding to the propagator new
phenomenolo\-gical pieces driven by positive mass dimension
parameters:~\footnote{Such ansatz parallels those make at zero
temperature in the presence of condensates~\cite{Chetyrkin:1998yr}.}
\begin{equation}
D_{00}^{NP}(\mathbf{k}) = \frac{m_G^2}{(\mathbf{k}^2+m_D^2)^2} \,.
\label{eq:np_prop}
\end{equation}
Adding the perturbative and non perturbative contributions, we get for
the Polyakov loop
\begin{equation}
-\log L = -\frac{N_c^2-1}{4N_c}\frac{g^2 m_D}{4\pi T} + \frac{g^2 \langle A_{0,a}^2\rangle_T^{NP}}{4N_c T^2} \,.
\label{eq:log_L}
\end{equation}

{\bf Comparison to lattice data.} The perturbative contribution to the
Polyakov loop has logarithmic corrections in $T$. We will assume that
$\langle A_{0,a}^2\rangle_T^{NP}$ is constant in $T$, or at most
logarithmic. This means that $\log L$ has a temperature power
correction.

\begin{figure}[tbp]
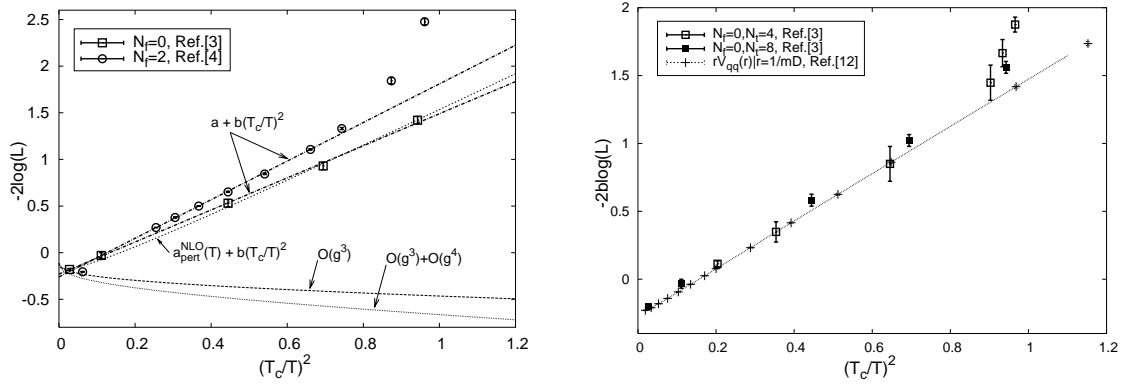

\begin{minipage}[t]{7.3cm}
\includegraphics[width=7.3cm]{qmfig1.epsi}
\end{minipage}
\hspace*{0.3cm}
\begin{minipage}[t]{7.3cm}
\includegraphics[width=7.3cm]{qmfig2.epsi}
\end{minipage}
\caption{The logarithmic dependence of the renormalized Polyakov loop
 versus the inverse tempera\-ture squared in units of the critical
 temperature. Lattice data from
 \cite{Kaczmarek:2002mc,Kaczmarek:2005ui}. At the left we plot fits
 with $a$ adjustable constant or predicted by NLO perturbation theory
 \cite{Gava:1981qd}. Purely perturbative LO and NLO results for
 $N_f=0$ are shown for comparison. The right figure illustrates the
 duality between the Polyakov loop and the quark-antiquark potential
 at zero temperature. The line represents $r V_{qq}(r)$, takes from
 lattice data~\cite{Necco:2001xg}, and modified with the
 change~$r=1/m_D$.}
\label{fig:log_plot}
\end{figure}

The lattice data of the renormalized Polyakov loop of
Ref.~\cite{Kaczmarek:2002mc} for $N_f=0$ flavors and
Ref.~\cite{Kaczmarek:2005ui} for $N_f=2$ flavors have been displayed
in Fig.~\ref{fig:log_plot}.
These data suggest a linear fit of the form $-2\log L = a + b \left( T_c/T \right)^2 $~\cite{Megias:2005ve,Megias:2004hj}, which yields
\begin{equation}
a = \left\{\matrix{
-0.23(1)  \,, && \!\!\!\!\!\!\!\!\!\! N_f=0   \cr
-0.26(1)  \,, && \!\!\!\!\!\!\!\!\!\! N_f=2  
} \right. \,,
\qquad
g^2\langle A_{0,a}^2\rangle_T^{NP} = \left\{\matrix{
(0.87 \pm 0.02 \; \textrm{GeV})^2 \,, && \!\!\!\!\!\!\!\!\!\! N_f=0  \cr 
(0.71 \pm 0.01 \; \textrm{GeV})^2 \,, && \!\!\!\!\!\!\!\!\!\! N_f=2
} \right.  \,,
\label{eq:fit_value}
\end{equation}
with $\chi^2/DOF=0.45,\,6.14$, for $N_f=0,\,2$ respectively. The
perturbative results for the highest temperature~$6T_c$ are in
qualitative agreement with the fitted values for $a$,
Eq.~(\ref{eq:fit_value}).~\footnote{We consider the LO and NLO
perturbative result of~Ref.~\cite{Gava:1981qd}} 

Finite temperature results for the pressure in pure
gluodynamic~\cite{Kajantie:2000iz} yield for the gluon con\-densate $(0.93(7)
\; \textrm{GeV})^2$ (in the temperature region used in our fits and in
Landau gauge). Our result is also in reasonable agreement with
existing zero temperature values~\cite{RuizArriola:2004en}.

\begin{figure}[tbp]
\includegraphics[height=5.5cm]{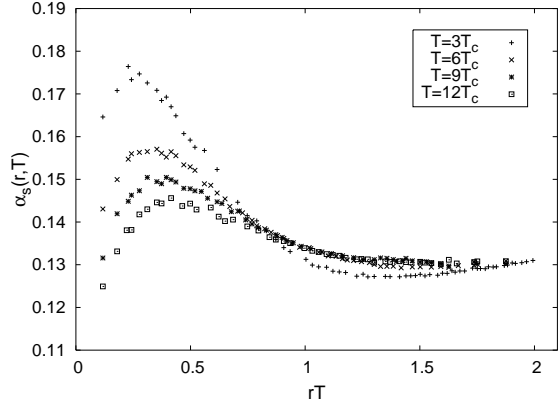}
\caption{ The $r$ dependence of the running coupling in pure
gluodynamic for different values of temperature. It is obtained by
mathing Eq.~(\ref{eq:free_energy}) to lattice data for the free
energy from Ref.~\cite{Kaczmarek:2004gv}.}
\label{fig:alphas}
\end{figure}

\newpage
{\bf Heavy quark free energy.}  The quark-antiquark potential can be
related to the scattering amplitude corresponding to one gluon
exchange. In the non-relativistic limit
\begin{equation}
F_1(\mathbf{x},T) = -\frac{4}{3} g^2 \int \frac{d^3 k}{(2\pi)^2} e^{i{\mathbf k} \cdot {\mathbf x}} D_{00}(\mathbf{k}) \,.
\end{equation}
We can consider non perturbative contributions for the free energy by
adding the non perturbative term~(\ref{eq:np_prop}) to the
perturbative propagator, i.e. $D_{00}(\mathbf{k}) =
D_{00}^P(\mathbf{k}) + D_{00}^{NP}(\mathbf{k})$. At LO, ${\cal
O}(g^2)$, and NLO, ${\cal O}(g^3)$, the singlet free energy has the
form
\begin{equation}
F_1(\mathbf{x},T) = -\frac{N_c^2-1}{2N_c} \left(
\frac{g^2}{4 \pi r} + \frac{1}{N_c^2-1}\frac{g^2 \langle A_{0,a}^2 \rangle_T^{NP}}{T}
\right) e^{-m_D r}
-\frac{N_c^2-1}{2N_c}\frac{g^2 m_D}{4\pi} + \frac{g^2 \langle A_{0,a}^2\rangle_T^{NP}}{2N_c T} \,.
\label{eq:free_energy}
\end{equation}
Taking the value obtained in (\ref{eq:fit_value}) for $g^2 \langle
A_{0,a}^2\rangle_T^{NP}$, we can match the free energy to lattice
data~\cite{Kaczmarek:2004gv}. This way, one obtains the $r$ and $T$
dependence of the running coupling~$\alpha_s$. This is displayed in
Fig.~(\ref{fig:alphas}). One can see a smooth behaviour, and it
confirms the better fit of the Polyakov loop with
$a=\textrm{constant}$. The values are relatively small, which is in
contrast with existing analysis of $\alpha_s$ at finite
T~\cite{Kaczmarek:2002mc,Kaczmarek:2004gv}.

In we consider the limit of large separation in the free energy,
Eq.~(\ref{eq:free_energy}), one reobtains $F_\infty (T) = -2T \log L
$. In the zero temperature limit (we take $m_D r \rightarrow 0$ in
(\ref{eq:free_energy}))
\begin{equation}
F_{1}({\mathbf x},T) \stackrel{T \rightarrow 0} \sim 
-\frac{N_c^2-1}{2N_c}\frac{g^2}{4\pi r} 
+ \frac{g^3 \langle A_{0,a}^2 \rangle_{T=0}^{NP}}{2N_c}r \equiv V_{qq}(r) \,,
\label{eq:t_zero}
\end{equation}
which is just the quark-antiquark potential at $T=0$~\cite{Necco:2001xg}. The
Coulomb term is the standard perturbative result at LO. The second one
is a non perturbative linear contri\-bution, and we get for the string
tension $\sigma = g^3 \langle A_{0,a}^2 \rangle_{T=0}^{NP}/2N_c$. From
Eqs.~(\ref{eq:log_L}) and (\ref{eq:t_zero}) we deduce a property:
\begin{equation}
F_\infty (T) = V_{qq}(r)|_{r=1/m_D} \,, 
\label{eq:duality}
\end{equation}
which is valid if we assume that $\alpha_s(r,T=0) \langle A_{0,a}^2
\rangle_{T=0}^{NP} = \alpha_s(r \rightarrow \infty, T) \langle
A_{0,a}^2 \rangle_T^{NP}$.  This duality between $F_\infty(T)$ and
$V_{qq}(r)$ is only valid at LO in perturbation theory.~\footnote{The
duality is formal, i.e. we are assuming $g = \textrm{constant} \ne
g(r,T)$. If we take into account the different assymptotic behaviours
of $\alpha_s$: $\alpha_s(r) \equiv \alpha_s(r,T=0)$ and $\alpha_s(T)
\equiv \alpha_s(r\rightarrow \infty, T)$; Eq.~(\ref{eq:duality})
writtes
\vspace{-0.3cm}
\begin{equation}
b F_\infty (T) = V_{qq}(r)|{r=\gamma/T} \,,
\quad \textrm{with} \quad 
b=(\alpha_s(r)/\alpha_s(T))^{3/4} 
\quad \textrm{and} \quad 
\gamma=(\alpha_s(r)/\alpha_s(T)^3)^{1/4} \,.
\end{equation}
}

To check numerically Eq.~(\ref{eq:duality}), we plot in
Fig.~\ref{fig:log_plot} the lattice data for $-2 b \log L$ versus
$(T_c/T)^2$. We can see a remarkable agreement. This duality looks
deeply into the analogy between the quark-antiquark potential at zero
temperature and the Polyakov loop.

\vspace{0.4cm} Work supported by funds provided by the Spanish DGI
Grant No. BFM2002-03218, Junta de Andaluc\'{\i}a Grant No. FM-225,
and EU RTN Contract No. CT-2002-0311.

\end{document}